\documentstyle[12pt,epsf]{article}
\def\reference{\parskip 0pt\par\noindent\hangindent 0.5 truecm}
\def\kms{km ${\rm s}^{-1}$}
\begin{document}
%
% Title
% Capitalise the title normally - do not use ALL CAPS.
%
\title{Optical Counterparts to Galaxies in the Cen A Group 
Discovered by HIPASS}
%

% Authors
% Here comes the author(s) of the paper. Please add the appropriate author
% names for your paper and indicate within the $^...$ the number(s)
% which corresponds to the institute(s) of each author. In this example
% the second author has two institutional affiliations.
% Add or remove authors as required, maintaining the \and syntax between
% each author, but no \and after the last author.
% **** IMPORTANT: Leave the closing curly bracket line as is. ******

\author{Patricia M. Knezek $^{1,2}$ 
} % IMPORTANT: leave this curly bracket as the first character of this line.

% Date - leave this blank.
\date{}
\maketitle

% Institutions
% Here fill in your institute name(s) and address(es)
% The number in $^...$ indicates the author number.  For example
{\center
$^1$ Department of Physics and Astronomy, The Johns Hopkins University, Bloomberg Center, Baltimore, MD, 21218-2695, USA\\pmk@pha.jhu.edu\\[3mm]
$^2$ Visiting Astronomer, Cerro Tololo Inter-American Observatory, operated by AURA, Inc.\ under contract to the National Science Foundation.
}

% Abstract
% Simply place your abstract between the \begin{abstract} and
% \end{abstract} commands.
%
\begin{abstract}
% Place the abstract here.

We have completed a 21-cm survey of 
a 600 square degree region of the
Centaurus A group of galaxies at a redshift of $\sim$500 \kms\ as part of
a larger survey of the entire southern sky. 
This group of galaxies was recently the subject of a
separate and thorough optical survey (C{\^o}t{\'e} et al.\ 1997),
and thus presented an ideal comparison for us to test the survey performance.
We have identified 10 new
group members to add to the 21 already known in our survey area. Six of the
new members are previously uncatalogued galaxies,
while four were catalogued but assumed not to be group members.
Including the 7 known members outside of our survey area, this brings
the total known number of Cen A members to 38.  
All of the new H{\sc i} detections have optical counterparts, most being
intrinsically very faint (M$_{B} >$ -13.0), late-type low surface brightness
dwarfs. 
Most of the new members have H{\sc i} masses only a few times 
our survey limit of
10$^{7}$ M$_{\odot}$ at an assumed distance for the group of 3.5 Mpc, 
and are extremely gas-rich, with a median M$_{HI}$/L$_B >$ 1. 
Our limiting H{\sc i} sensitivity
was actually slightly worse than the H{\sc i} follow-up observations of the
C{\^o}t{\'e} et al.\ optical survey, yet {\it we have already increased the
known number of group members by 50\% using an  {\rm H}{\sc i} survey technique.}
While we have increased the known {\it number} of members by $\sim$50\%,
these new members 
contribute $<$4\% to its light. 

\end{abstract}

{\bf Keywords:}
% Place keywords here.  PASA uses the standard list of subject 
% headings adopted by The Astrophysical Journal and available from URL:
%   http://www.noao.edu/apj/keywords96.html
galaxies: general - galaxies: luminosity function, mass function -
galaxies: statistics

% A formatting command to add space between the author list and the body
% of the paper when printed. This spacing may be changed as desired.
\bigskip

%
% Body of paper
%

\section{Introduction}

Most galaxies are thought to live in physical groups with a few
to perhaps one hundred gravitationally bound members.
Typically, these groups of galaxies are dominated by one to a few
massive spirals and/or ellipticals which are surrounded by less
massive ``dwarfs''.  Indeed, our own Local Group is now thought to contain 
38$^{+6}_{-2}$
members, and is dominated by the Milky Way and M~31
(Mateo 1998) both in terms of mass and luminosity.  Yet only $\sim$10\% of
the Local Group {\it by number} have $M_B < -18$.  The clues
to the formation and evolution of the Local Group - including the
Milky Way - are closely tied to understanding the {\it dwarf}
galaxy population.  Furthermore, understanding where the Local Group fits in
the formation and evolution of the universe requires determining if
the Local Group is truly an ``average'' group of galaxies.
Does 90\% of the population of most groups galaxies have
$M_B > -18$?  What fraction of the dwarf population is due to dwarf
ellipticals and spheroidals (dE/dSphs), and what fraction is due to
dwarf irregulars and dwarf spirals (dIrr/dSps)?  Is there an environmental
difference in this fraction depending on the morphological type of the
gravitationally dominant members, or perhaps the depth of the potential
well?  Are most groups of galaxies truly bound systems?  Are they
typically in virial equilibrium?  Answers to all of these questions depend on
determining the complete membership of individual groups, which implies
both needing to identify the members (many of which may be intrinsically
quite faint) {\it and} determining their distance.

One of the nearest group of galaxies is the Centaurus A group (Cen A group).  
This group of galaxies not only contains NGC~5128 (Centaurus A), the 
nearest known radio galaxy, but according to de Vaucouleurs (1979), it
has the largest spread of morphological types of any nearby group of
galaxies.  All of the most luminous members appear disturbed, leading Graham
(1979) and van Gorkom et al.\ (1990) to suggest that perhaps this group
has recently accreted a population of gas-rich dwarfs.   C\^ot\'e et al.\
(1997) recently completed a study of this group.  They optically searched for
potential members using the SRC $J$ survey films, then confirmed 
membership by follow-up H$\alpha$ and H{\sc i} observations.  
They searched over a total area of approximately 900 square degrees, 
and in all, identified 27 group members.  An additional member was discovered
by Matthews \& Gallagher (1996), bringing the total number of accepted 
members to 28 prior to the current study.  

Given the recent, thorough, {\it optically based} study of the Cen A
group, along with its intrinsic scientific interest, the Cen A group was
selected as an ideal location to begin the H{\sc i} Parkes All Sky Survey
(H{\sc i}PASS), now being conducted on the CSIRO Parkes 64~m radio telescope in
Parkes, Australia.  The assumed distance of the Cen A group (3.5 Mpc,
based on Cepheid distances for group members
NGC~5253, Sandage et al.\ 1994, and NGC~5128, Hui
et al.\ 1993) is
such that H{\sc i}PASS should detect galaxies with H{\sc i} masses of
$> 10^7$ M$_{\odot}$, which compares very favorably with the limiting
sensitivity of the H{\sc i} follow-up studies of C\^ot\'e et al.\ (1997).
How does the population of group members identified via an H{\sc i} search
compare with the population identified through a deep optical search?
Of the 28 identified Cen A group members, 
21 lie within the
600 square degrees of this initial H{\sc i}PASS survey, where the 
boundaries were selected
to fit in with the scanning grid of the
larger  H{\sc i}PASS survey.  The C\^ot\'e survey region was
12$^{h}$ 30$^{m}$ $< \alpha <$ 15$^{h}$; -20$^{\circ} > \delta >$ -50$^{\circ}$.
This extends further north than the H{\sc i}PASS region discussed here.
A complete discussion of the H{\sc i} properties of detected members of the
Cen A group will be presented in a separate paper (Banks et al.\ 1999).  Here
we concentrate on the first results of the {\it optical} follow-up of
the H{\sc i}PASS observations.

% Place contents of first section here.

\section{Observations}

\subsection{H{\sc i} Observations}

The H{\sc i}PASS, as mentioned above, is being conducted on
the 64~m radio telescope at Parkes, Australia. It uses the Multibeam system,
which comprises 13 separate beams (26 receivers).  This system surveys 
1024 velocity channels simultaneously as the telescope is scanned across the sky
in declination at a rate of 1 degree/minute.  For H{\sc i}PASS the velocity 
coverage is -1200 to 12700 \kms.  Scans are stored every 5 seconds, and
the average system temperature is 23~K.  
In order to allow for rejection of time-dependent interference, 
the scanning pattern has been 
designed such that every source passes through several beams during a scan, 
and through several different scans separated by days or weeks of time. 
The effective total integration time per beam is 450 seconds, and 
the 5$\sigma$ sensitivity is measured to be 70 mJy/channel/beam.

The channel spacing is 13.2 \kms\ per channel, but the
spectra are smoothed on-line using a 25\% Tukey filter, to
reduce  ``ringing'' caused by the strong Galactic signal entering
through spectral sidelobes. This causes a loss in velocity resolution
of 37\%. Consequently velocity resolution is 18.2 \kms,
or 35 \kms\ after  Hanning smoothing. The on-line
data reduction, including the bandpass correction, is
described in Barnes et al.\ (1998), while the whole Multibeam system
is described in Staveley-Smith et al.\ (1996).
H{\sc i}PASS started early in 1997, and the southern sky should be complete
within 3 years.

Data cubes are generated from the observations by gridding the individual
spectra with 4$^{\prime}{\rm x}4^{\prime}$ pixels.  A by-eye search of
velocity space from 0 to 1000 \kms\ resulted in many detections below
200 \kms, most of which were assumed to be due to high velocity clouds
within our own Galaxy (Wakker 1991).  
Thus we chose not to include those detections,
with the exception of one source at v$_{\rm hel}=122$ \kms, which has
previously been identified as a member of the Cen A group (CEN~5, 
C\^ot\'e et al.\ 1997).  Including CEN~5, 28 clear detections were found
in the search, of which 18 were previously known members.  The three
known members within our search area which were {\it not} detected by
H{\sc i}PASS are Centaurus A itself, which is strongly self-absorbing
at 21 cm, and NGC~5206 and ESO~272-G025, neither of which were detected
in H{\sc i} 
by C\^ot\'e et al.\ (1997) at a limit of $< 5.5{\rm x}10^6$ M$_{\odot}$.
Thus we have recovered all of the known members with H{\sc i} masses
above our sensitivity limit, and in addition we have identified 10 new
candidate group members.

\begin{figure}
\begin{center}
\leavevmode
\epsfysize=250pt \epsfbox{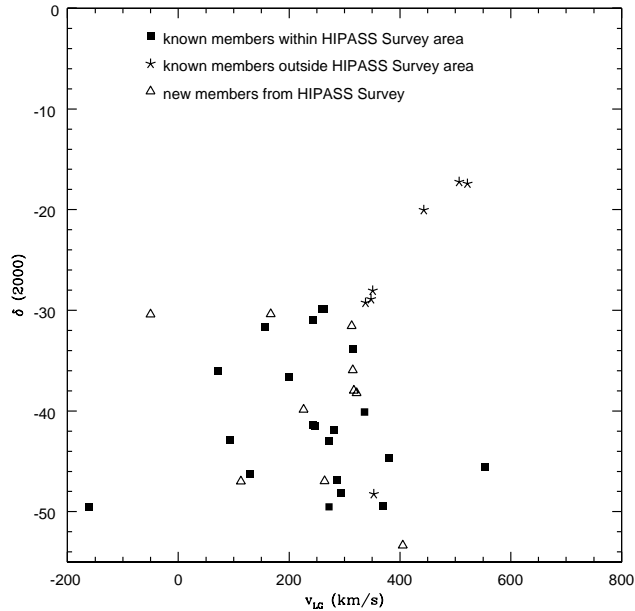}
\caption{A graph of the velocities of the known Cen A group members,
corrected relative to the velocity of the Local Group, versus declination.
}
\label{vlg}            % for cross-references
\end{center}
\end{figure}

The new candidate members of the Cen A group are found scattered
through
the known distribution of velocities for the group, as can be seen in
Figure~\ref{vlg}.  For this figure, we have corrected the velocities of the
galaxies to their velocity with respect to the Local Group, v$_{\rm LG}$,
following the prescription of Yahil et al.\ (1977).  The filled squares
represent previously identified galaxies within the H{\sc i}PASS survey
area, the stars represent previously identified galaxies outside the
H{\sc i}PASS survey area, and the open triangles represent new candidate
group members identified through the H{\sc i}PASS survey. The
lack of new group members north of $\delta=-30^{\circ}$ is artifical,
since that region was not included within the original 600 square
degrees surveyed in this study.  The apparent cutoff in velocity at
v$_{\rm LG} \sim 400$ km s$^{-1}$, however, appears real.  Scanning the
data cubes to v$_{\rm hel}=1000$ \kms\ should detect systems with 
v$_{\rm LG} < 735$ \kms\ in the direction of the Cen A group. All of the
newly detected galaxies had H{\sc i} masses several times higher than the
nominal sensitivity limit of the H{\sc i}PASS survey, thus it is unlikely
that candidate members were missed at higher velocities because their
H{\sc i} masses fell below the sensitivity limit due to their presumably
being slightly farther away.

\subsection{Optical Observations}

$B$ and $R$ images were obtained of most of the candidate group members without 
previous photometry using either the 40-inch Swope telescope at Las Campanas
Observatory (LCO) in Chile or the 40-inch telescope at 
Siding Springs Observatory (SSO) in Australia.  
The two primary observing runs were at LCO on 10--16 April 1997 and
on 23--28 July 1997.
Those galaxies observed in non-photometric conditions were calibrated 
with observations at SSO during January and April 1998, with a few additional
calibrations using the Curtis Schmidt telescope at CTIO in April 1998.
A $2048{\rm x}2048$ chip was used, except during the 
January 1998 run at SSO, when a $800{\rm x}800$ chip was used. 
For the two primary observing runs, the Tek\#1 CCD chip was used, 
which is a thinned, blue-sensitive
CCD.  The field of view was 24$^{\prime}$ with the chip mounted 
at f7.5, and  pixel size was 0.7$^{''}$.  
Typical total integrations times are 3600 seconds in $B$, and 1800 seconds 
in $R$.  Only the $B$ images have been used for the following analysis.
For the galaxy H{\sc i}PASS~1337-39, the blue magnitude has been estimated
from the Digitized Sky Survey, and should be considered to be very 
uncertain.

\begin{figure}
\begin{center}
\leavevmode
\epsfbox{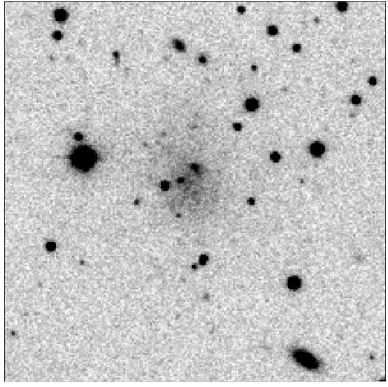}
\caption{A $B$ image of one of the newly discovered Cen A dwarfs, 
H{\sc i}PASS1321-31.  North is
up and east is to the left. The image is $\sim$2.7$^{\prime}$ across.}
\label{hp1321b}            % for cross-references
\end{center}
\end{figure}

Data processing proceeded in the usual manner.
The data were overscan-subtracted, and then bias subtracted 
using a median filtered bias from each night.  Primary flat fielding was
done using sky flats obtained each evening.
Several Landolt standard fields (Landolt 1992) were observed each night in
the broadband filters. 
The atmospheric extinction was determined
using the Landolt standards and found to be similar to the standard 
CTIO extinction coefficients.
Figure~\ref{hp1321b} shows an example of the final $B$ image of 
one of the newly identified galaxies, H{\sc i}PASS1321-31.
Optical counterparts were identified for {\it all} of the new Cen A
group members detected by H{\sc i}PASS.  

\section{Discussion}

The galaxy presented in Figure~\ref{hp1321b}, H{\sc i}PASS1321-31, 
is typical of the
dwarfs which were discovered through the H{\sc i}PASS.  In general
these dwarfs are very low surface brightness ($\mu_0=24.2$ mag 
arcsec$^{-2}$ in $B$ for H{\sc i}PASS1321-31), with ``by-eye'' 
diameters of order
30$^{\prime\prime}$, which corresponds to $\sim$0.5 kpc at the distance
on the Cen A group.  This is comparable to the values found for Local
Group dwarf galaxies, which have diameters between 0.18 and 3 kpc, based
on their disk scale lengths (Mateo 1998).  

The newly detected dwarfs
are also represented by H{\sc i}PASS1321-31 morphologically.  Like 
H{\sc i}PASS1321-31, most of the new members appear to be dwarf
irregulars.  This is not unexpected, as the H{\sc i}PASS is
going to select out galaxies which possess atomic gas, introducing a bias
toward late-type systems.  
A few do have higher surface brightness inner regions, 
similar to some of the late-type galaxies with surface brightness ``steps'' 
discussed by Matthews \&
Gallagher (1997).  Also similar to
the Matthews \& Gallagher systems, a few appear to have unresolved,
point-like nuclei.  In this respect, they differ from Local Group
dwarfs, where only gas-poor, dwarf spheroidals, have nuclei.  
Apparently, even these small systems can have
multi-component disks and/or a nuclear component, 
suggesting a complex evolutionary history.

The Cen A group also appears to differ from the Local Group, and 
other groups of galaxies, in the morphological composition of the
group.  It has already been noted by de Vaucouleurs (1979) that
the Cen A group has the largest morphological variety seen in any
nearby group.  The dwarf population only has two members which
C{\^o}t{\'e} et al.\ (1997) claim can be classified as early-type
dwarfs, NGC~5206 and NGC~5237, based strictly on optical morphology.
Although detailed morphological classification has not yet been
possible, the optical appearance of all the newly detected dwarfs
is consistent with their being late-type dwarfs.  Thus, while at
least 30\% of the Local Group members with M$_{\rm B} < -11$ appear
to be early-type systems, only $\sim$7\% of the Cen A
group members currently known are early-type, unless the two 
luminous S0s are included, NGC~5102 and NGC~5128.  If the two
groups actually have a similar morphological composition, this
implies that the Cen A group contains at least 60 members.

\begin{figure}
\begin{center}
\leavevmode
\epsfysize=250pt \epsfbox{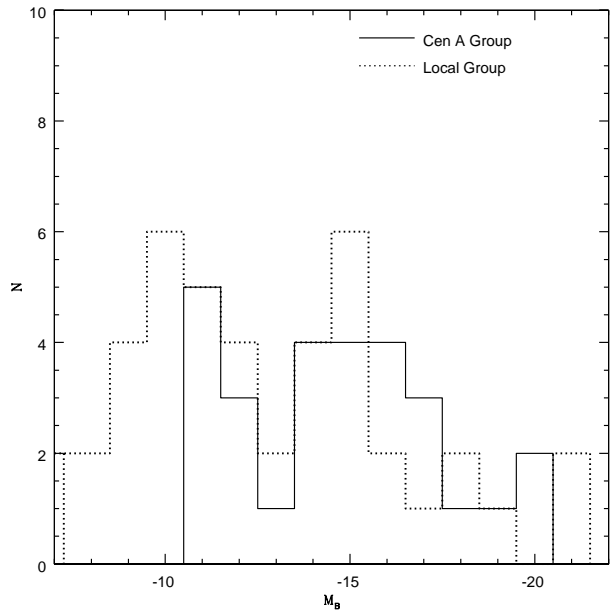}
\caption{A histogram of the absolute blue magnitudes of the known Cen A
group members within the H{\sc i}PASS survey area and 
the known Local Group members.
}
\label{bmags}            % for cross-references
\end{center}
\end{figure}

In order to study the optical integrated properties of the Cen A group,
we present a histogram of 
the distribution in absolute $B$ magnitudes in Figure~\ref{bmags} for 
the members of the Cen A group {\it within the Survey area} (solid line)
with $B$ photometry, in comparison with a histogram of all known Local
Group members (dotted line).  We have chosen to display only the 
Cen A group members within our survey area since the correction for
dwarfs outside the survey area is unknown.  The absolute magnitudes for
the Cen A group members were determined assuming a distance of 3.5 Mpc,
and corrected for Galactic extinction.  Internal extinction corrections
have not been made.  The Local Group data are from Mateo (1998), except
for M~31, M~33, the LMC, the SMC, which are from de Vaucouleurs et al.\ 
(1991), scaled to
the distances given in Mateo (1998), and the Milky Way, which is from
van den Bergh (1992).

The dearth of galaxies in the Cen A group seen in 
Figure~\ref{bmags} at M$_{\rm B} > -11$ is a
selection effect.  Most of the Local Group galaxies that are that faint are
dwarf spheroidals with no detected H{\sc i} gas.  The three late-type
Local Group 
members which do have such low luminosities have M$_{\rm HI}$/L$_{\rm B}$
of $\sim$ 0.2 - 1.4.  H{\sc i}PASS would not detect galaxies that faint
unless they had an ratio of M$_{\rm HI}$/L$_{\rm B} > 7$, while 
the C\^ot\'e et al.\ (1997) study deliberately 
targeted late-type dwarf galaxies.  Furthermore, at the distance of the
Cen A group, a M$_{\rm B} \sim -10$ implies an integrated apparent magnitude
of m$_{\rm B} > 17.7$ {\it not} including Galactic extinction, which is 
generally significant for this group of galaxies (values for A$_{\rm B}$ 
range from 0.14 - 2.12, with $\sim$ 1/3 of the galaxies having 
A$_{\rm B} > 0.50)$.  Thus, the determination
of membership at the faint end of the luminosity function will have to
wait for more sensitive studies.  Even though the newly detected galaxies
all have M$_{\rm B} < -10.5$, however, they contribute very little to the
integrated blue luminosity of the group.  After correcting for Galactic
extinction, the new members are found to add $<4$\% to the {\it total}
blue luminosity of the group.

Overall, the two groups have similar luminosity distributions, at least to 
M$_{\rm B} \sim -11$.  This is consistent with 
the luminosity function being the same in both groups.  For the Local 
Group, Mateo (1998) found the best fitting Schechter (1976) function to
have $\alpha = -1.16$, and M$_{\rm B} = -21.42$.  It is interesting that
the luminosity distribution of the two groups is so similar, while the
morphological composition is so different.  As noted above, for  
M$_{\rm B} < -11$, 
about 30\% of the Local Group are early-type galaxies, whereas $<$15\%
of the Cen A group appear to be early-types, and the percentage is only
that high if you include the luminous S0s NGC~5102 and NGC~5128.  If
the H{\sc i }PASS and C{\^o}t{\'e} et al.\ (1997) survey techniques have
truly missed the corresponding early-type, gas poor galaxies in the
Cen A group, then the luminosity functions of the two groups may differ
significantly.  If the luminosity functions are actually the same, then,
at least for M$_{\rm B} < -11$, {\it the morphological composition of
the two groups differs significantly}.

\begin{figure}
\begin{center}
\leavevmode
\epsfysize=250pt \epsfbox{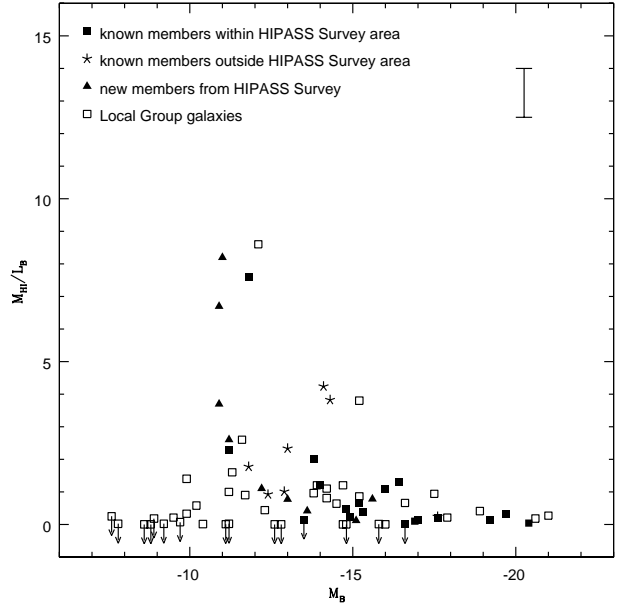}
\caption{A graph of M$_{\rm HI}$/L$_{\rm B}$ versus M$_{\rm B}$ 
of the known Cen A group members and the Local Group galaxies. The errorbar
indicates a typical uncertainty for the {\it new} galaxies detected in this
study.  The uncertainty is smaller for the previously known galaxies.}
\label{mhlb}            % for cross-references
\end{center}
\end{figure}

Another way of stating the difference between the two groups is to look
at the ratio of gas mass to blue luminosity for the individual
members.  
The ratio of atomic gas mass-to-blue luminosity, M$_{\rm HI}$/L$_{\rm B}$,
is shown in Figure~\ref{mhlb} versus the absolute blue magnitude, M$_{\rm B}$,
for all Cen A group members for which optical photometry was available, and
for the Local Group.  The H{\sc i} mass of NGC~5128 is from Richter
et al. (1994).  Local Group H{\sc i} data are from Mateo (1998), except
for M~31, M~33, the LMC, the SMC, and the Milky Way, which are from Lang (1980),
scaled to the distances given in Mateo (1998).
The error bar shown in the upper right corner of Figure~\ref{mhlb} is a
representative
error bar for the {\it new} galaxies identified in this study, assuming
a magnitude error of 0.3 mags and an H{\sc i} mass error of 50\%.  This
is very likely an overestimate of the errors.  The mean 
M$_{\rm HI}$/L$_{\rm B}$ for these new galaxies is 2.7, while the median
is $\sim$1.5.  Thus, these galaxies have typical atomic gas to blue
light ratios which are $\sim$10 times higher than that found in late-type
disk galaxies, and which are comparable to the highest values found
by Matthews \& Gallagher (1997) in their study of extreme late-type
galaxies.

From Figure~\ref{mhlb} it can be seen that the ratio of gas mass to blue
luminosity is similar for the brightest galaxies in both the Cen A group
and the Local Group.  However, for $-16 < {\rm M_{B}} < -11$,  a significant
number of the galaxies in the Cen A group appear to be gas rich.   In fact,
more than three times as many galaxies in the Cen A group have 
M$_{\rm HI}$/L$_{\rm B} > 2$ as in the Local Group.  If Cen A has 
recently acquired a population of dwarfs which have not yet been
significantly affected by the group dynamics, as suggested by Graham 
(1979) and van Gorkom et al.\ (1990), this might explain the population
of extremely gas-rich dwarfs.  However, the 
line-of-sight velocity dispersion of the
group is only $\sim$135 \kms\ within the H{\sc i}PASS survey region, and
only $\sim$140 \kms\ including all known members.  As noted by 
C{\^o}t{\'e} et al.\ (1997), while this implies that the Cen A group is
not virialized, the velocity dispersion is quite typical of groups of
galaxies of this size.  It is not obvious, then, that this group of 
galaxies has accreted a new, gas rich population of dwarfs, while other
similar groups of galaxies apparently have not.

\section{Conclusions}

Our initial survey of 600 square degrees of the Centaurus A group of
galaxies in 21-cm has yielded 10 new candidate members, bringing the
number of known members within this initial H{\sc i}PASS survey 
region to 31, and the total
number group members to 38.  The new members add $\sim$50\% to the 
number of members known, but only $\sim$4\% to the total luminosity
of the group.  The new members all have optical
counterparts, and most are very faint, late-type, gas-rich 
low surface brightness
dwarfs.  Due to our survey technique, and that of the previous survey
by C{\^o}t{\'e} et al.\ (1997), very little is known about the possible
early-type (dE/dSph) population of the Cen A group.  However, the 
luminosity {\it distribution} of the Cen A galaxies is entirely consistent
with the Local Group luminosity distribution to the optical limit of
galaxies detected within the Cen A group.  While 30\% of the Local Group
galaxies with M$_{\rm B} < -11$ are early-type systems, $<$15\% of the
known Cen A galaxies are.  If there is a Local Group type 
population of early-type
galaxies, then the luminosity distributions of the two groups may be
significantly different.

% Place contents of next section here.

%
% Add as many section titles/contents as required.
%
% If you have subsections then use the
% \subsection{SUBSECTION TITLE}
% command and if you have subsubsections then use the
% \subsubsection{SUBSUBSECTION TITLE}
% command.  To use these commands, 
% first remove the % from the start of the line.

% It is preferable to embed your figures in the text. 
% One way to do this is to use the psfig style file and use the following
% commands to include the figures:

% \begin{figure}
% \begin{center}
% \psfig{file=filename.ps,height=10cm}
% \caption{Write your figure caption here.}
% \label{figlabel}            % for cross-references
% \end{center}
% \end{figure}

% To use the above commands, first remove the % from the beginning of
% the lines and then fill in your own values etc as appropriate.

% Tables
% Please consult previous issues of PASA
%  to see how tables are to be formatted.

\section*{Acknowledgements}
I would like to thank Gareth Banks, Mike Disney, and Robert Minchin for
much of the data reduction and analysis that is used in this paper,
as well as the entire Multibeam Working Group.  This research is based
on results for the H{\sc i}PASS being conducted using the CSIRO Parkes
telescope in Parkes, Australia.  The Johns Hopkins
University has generously supported my frequent observing trips to
CTIO.  This research has made use of the Digitized Sky Survey,
produced at the Space Telescope Science Institute under U.S.\ Government 
grant NAG W-216.  
It has also made use of the NASA/IPAC Extragalactic Database (NED) 
which is operated by the Jet Propulsion Laboratory, California Institute of 
Technology, under contract with the National Aeronautics and Space 
Administration. 

% Place acknowledgements here. Omit above \section command if there
% are no acknowledgements.

\section*{References}

% PASA uses the same conventions as ApJ for journal abbreviations.  Sample
% references are as follows. 
% Please follow the same format for your references.

%\reference Author, A.B. 1990 PASA 7, 2, 350

% for a journal article, or

% \reference Author, A.B. 1990 in This Is A Book Title, ed. Editor, C.D.,
% This Is A Publishers Name, 437

% for a book.

\reference Banks, G.\ et al.\ 1999, ApJ, in press
\reference Barnes, D.\ G., Staveley-Smith, L., Ye, T. \& Oosterloo, T., 1998, in Astronomical Data Analysis Software and Systems VII, eds. R. Albrecht, R.N. Hook\& H.A. Bushouse (San Francisco: ASP), 145, 89
\reference C\^ot\'e, S., Freeman K.\ C., Carnigan, C., \& Quinn, P.\  1997, AJ, 114, 1313
\reference de Vaucouleurs, G.\ 1979, AJ, 84, 1270
\reference de Vaucouleurs, G.\ et al.\ 1991, Third Reference Catalogue of Bright Galaxies (New York: Springer)
\reference Graham, J.\ A.\ 1979, ApJ, 232, 60
\reference Hui, X.\ et al.\ 1993, ApJ, 414, 463
\reference Landolt, A.\ U.\ 1992, AJ, 104, 340
\reference Lang, K.\ R.\ 1980, Astrophysical Formulae (Berlin: Springer-Verlag)
\reference Mateo, M.\ L.\ 1998, ARAA, 36, 435
\reference Matthews, L.\ \& Gallagher, J., 1997, AJ, 114, 1899
\reference Matthews, L.\ \& Gallagher, J., 1996, AJ, 111, 1098
\reference Richter, O.-G., Sackett, P.\ D., \& Sparke, L.\ S.\ 1994, AJ, 107, 99
\reference Sandage, A.\ et al.\ 1994, ApJ, 425, 14
\reference Schechter, P.\ 1976, ApJ, 203, 297
\reference Staveley-Smith, L., et al. 1996, PASA, 13, 243
\reference van den Bergh, S. 1992, A\&A, 264, 75
\reference van Gorkom, J.\ H., van der Hulst, J.\ M., Haschick, A.\ D., \& Tubbs, A.\ D.  1990,  AJ, 99, 1781
\reference Wakker B.\ 1991, A\&AS, 90, 495
\reference Yahil, A., Tammann, G., \& Sandage, A.\ 1977, ApJ, 217, 903

% Add as many references as required.

\end{document}